\documentclass[aps,prb,superscriptaddress,twocolumn,floatfix,showpacs]{revtex4}
\usepackage{amsmath}
\usepackage{graphics}

\newcommand{\vect}[1]{\boldsymbol{#1}}

\begin{document}

\title{Voltage-tunable singlet-triplet transition in lateral quantum
  dots}

\author{Jordan Kyriakidis}
\email{jordan@mailaps.org}
\affiliation{Department of Physics and Atmospheric Science, Dalhousie
  University, Halifax, Nova Scotia\ \ B3H 3J5, Canada}

\author{M. Pioro-Ladriere}
\affiliation{Institute for Microstructural Sciences, National Research
  Council of Canada, Ottawa, Ontario K1A 0R6, Canada}
\affiliation{CERPEMA, Universit\'{e} de Sherbrooke, Sherbrooke,
  Qu\'{e}bec J1K 2R1, Canada}

\author{M. Ciorga}
\affiliation{Institute for Microstructural Sciences, National Research
  Council of Canada, Ottawa, Ontario K1A 0R6, Canada}

\author{A. S. Sachrajda}
\affiliation{Institute for Microstructural Sciences, National Research
  Council of Canada, Ottawa, Ontario K1A 0R6, Canada}

\author{P. Hawrylak}
\affiliation{Institute for Microstructural Sciences, National Research
  Council of Canada, Ottawa, Ontario K1A 0R6, Canada}

\date{\today}

\begin{abstract}
  Results of calculations and high source-drain transport measurements
  are presented which demonstrate voltage-tunable entanglement of
  electron pairs in lateral quantum dots.  At a fixed magnetic field,
  the application of a judiciously-chosen gate voltage alters the
  ground-state of an electron pair from an entagled spin singlet to a
  spin triplet.
\end{abstract}

\pacs{73.21.La, 73.63.Kv, 85.35.Be, 03.67.Lx}

\maketitle

\section{Introduction}
\label{sec:introduction}

Proposals for spin-based quantum computation in a solid-state
environment\cite{baren95:condit.quant.dynam,brum97:coupl.quant.dots,loss98:quant.comput.quant.dots,kane98:silic.based.nuclear,vrijen00:elect.spin.reson}
require efficient techniques for manipulating the entanglement of
coupled qubits.
In this paper, we demonstrate theoretically and verify experimentally
that ground-state entaglement can be induced
\emph{solely} by applying a potential to the gates.  This is possible
because the gate voltage controls not only the chemical potential of
the dot, but the shape of the confining potential as well.
Consequently, the gate voltage can induce transitions in the dot
containing a well-defined, constant number of particles.

Early far-infrared measurements on arrays of few-electron
dots,\cite{sikor89:spect.elect.states,meurer92:singl.elect.charg} and
transport through single devices\cite{su92:obser.singl.elect} were
focused on the tunability of the electron number and, although
transitions were observed, it was difficult to assign, for example,
quantum numbers to these transitions.  Later experiments using
single-particle capacitance\cite{ashoor93:n.elect.groun} and
magneto-tunneling\cite{schmid95:quant.dot.groun} spectroscopy focused
on the evolution of the ground states as a function of magnetic field
and were able to distinguish features consistent with the two-electron
singlet-triplet transition.  High source-drain tunneling spectroscopy
probes the excited states as well as the ground state and therefore
the singlet-triplet transition can be more clearly and unambiguously
observed.  This has already been successfully applied to etched
vertical quantum dots with (spin) unpolarized
leads.\cite{kouwen97:excit.spect.circul,wiel98:singl.tripl.trans} In
the \emph{lateral} devices employed in the present study, the dot is
formed within a 2-dimensional electron gas (2DEG), with the lateral
confinement produced electrostatically by voltages applied to gates
located above the 2DEG.  Recent work employing a novel gate
design\cite{ciorg00:addit.spect.later} has allowed the electron-number
$N$ to be tuned down to a single electron.

There are at least two features unique in the lateral devices.  First,
since the leads are essentially 2DEG edges, applying a rather weak
magnetic field---approximately 0.4~T in practice---is sufficient to
produce spin-resolved edges.\cite{sachr01:spin.polar.injec} Therefore,
the tunneling rates into and out of the dots are significantly
different for each species of spin.  This spin-polarized injection
(and detection) allows us to distinguish orbital effects from spin
effects in transport
measurements.\cite{ciorg00:addit.spect.later,sachr01:spin.polar.injec,ciorg01:readout.singl.elect,ciorg02:collap.spin.singl}
Second, since the confinement potential is formed electrostatically by
the various gate voltages, altering the shape---in particular the
non-parabolicity---of the quantum dot can be accomplished while
keeping the particle number fixed.  It is important to note that the
familiar singlet-triplet transition is not caused by the difference in
Zeeman energy but rather by changes in the orbital part of the
wavefunction.  In multiple-dot systems, in particular with regards to
quantum-dot-based quantum
computation,\cite{loss98:quant.comput.quant.dots} this versatility is
of crucial importance.  Quantum-state engineering of this sort is
clearly observable in the experimental results we present below.

From the theoretical side, numerical analyses of the interacting
two-electron\cite{wagner92:spin.singl.spin,pfann93:compar.hartr.hartr,hawry93:singl.elect.capac}
problem, as well as higher electron
numbers,\cite{hawry93:singl.elect.capac,peder00:diffus.monte.carlo}
demonstrated singlet-triplet transitions in parabolic potentials using
either a fixed or an $N$-dependent harmonic frequency.  These works
focused on magnetic-field-induced transitions.  In our experiment,
the confining potential is a function of the \emph{continuous}
variables $V_g$ (the gate voltages) and the confinement in our
particular quantum dots deviate from parabolicity.  These features,
which are addressed in our theory, allow a spin phase diagram to be
constructed in the gate-voltage/magnetic-field plane and clearly
indicate how the $N=2$ singlet-triplet transition can be externally
engineered at fixed magnetic field.  In double-dot systems containing
one electron apiece, similar transitions, and for similar reasons,
have been theoretically demonstrated in both
lateral\cite{burkar99:coupl.quant.dots,hu00:hilber.space.struc} and
vertical\cite{burkar00:spin.inter.switc} devices.  Our theory is
applicable to these systems with only a few modifications related to
the orbital degrees of freedom---the spin physics are essentially
equivalent.

In the Coulomb-blockade regime, transport experiments probe the
two-electron system either by adding an electron to the one-electron
droplet, or by removing an electron from the three-electron droplet.
Each case corresponds to a distinct gate voltage, and in each case the
ground and excited states can be probed by high source-drain
spectroscopy, which directly reveals the singlet-triplet transition.
Our theory and experiment show that, for these two different gate
voltages, the transition between the entangled spin singlet
$\left|\uparrow\downarrow\right\rangle -
\left|\downarrow\uparrow\right\rangle$ and the spin triplet
$\left|\downarrow\downarrow\right\rangle$ occurs at two different
magnetic fields.

The paper is organized as follows:
Section~\ref{sec:experimental-results} describes the experimental
results, including the demonstration of a gate-voltage-induced
singlet-triplet transition.  Section~\ref{sec:theoretical-results}
contains a theoretical analysis, culminating in the spin phase diagram
of the $N=2$ interacting system in the gate-voltage/magnetic-field
plane.  Finally, Sec.~\ref{sec:conclusion} contains a concluding
discussion.

\section{Experimental results}
\label{sec:experimental-results}

\begin{figure}
  \resizebox{5.5cm}{!}{\includegraphics*{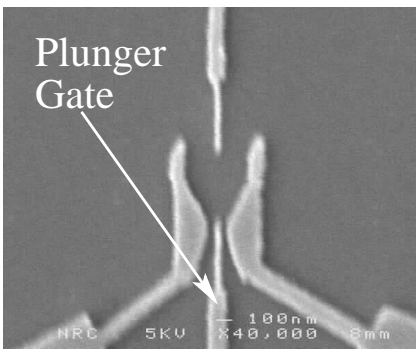}}
  \caption{SEM image of the gate geometry forming
    the quantum dot. This geometry enables a precisely known number of
    electrons ($N=0, 1, 2, \ldots, 50$) to be
    trapped\protect\cite{ciorg00:addit.spect.later} and produces a
    quasi-parabolic confinement potential.  Sweeping the plunger-gate
    voltage tunes both the shape and the chemical potential of the
    quantum dot.}
  \label{fig:gates}
\end{figure}

\begin{figure}
  \resizebox{8cm}{!}{\includegraphics*{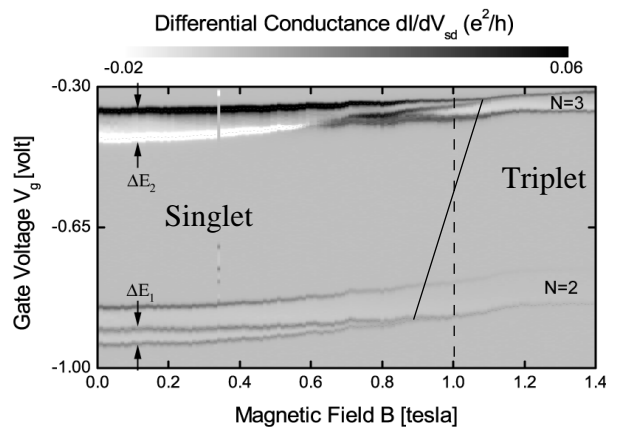}}
  \caption{High source-drain transport spectroscopy of the
    two-electron droplet.  The lower set of curves corresponds to
    fluctuations between $N=1$ and $N=2$, while the upper set
    corresponds to fluctuations between $N=2$ and $N=3$.  Both sets of
    curves probe the same states of the two-electron droplet (ground
    plus first excited state) but at different gate voltages.  The
    singlet-triplet transition is seen to occur at two different
    critical fields.  The solid line marks the singlet-triplet
    ground-state boundary.  This boundary can be traversed along the
    dashed line at $B=1$~T solely by adjusting the (plunger) gate
    voltage.}
  \label{fig:exp.sing-trip}
\end{figure}

An SEM image of a device similar to the one used in our experiments is
shown in Fig.~\ref{fig:gates}.
This geometry allows us to controllably tune the number of trapped
electrons in the 2DEG---90~nm below the surface---from about fifty
down to a single electron.\cite{ciorg00:addit.spect.later} High
source-drain transport measurements in the Coulomb-blockade regime
were carried out in order to detect both the ground and excited states
of the two-electron system.  Standard low-power ac measurement
techniques were used with a 10~$\mu$V excitation voltage applied
across the sample at a frequency of 23~Hz.  An additional dc voltage
was applied in order to obtain a high source-drain bias.  The
differential conductance $dI/dV_{\text{sd}}$ is measured directly in
such a configuration and the relevant data is shown in
Fig.~\ref{fig:exp.sing-trip} for a source-drain voltage of 350~$\mu$V.
In this figure we show the inverted grayscale for the $N=2$ subspace
as a function of magnetic field $B$ and plunger gate voltage $V_g$.
The negative differential conductance, which is also
tunable,\cite{ciorg01:tunab.negat.differ} is related to the
spin-polarized injection of electrons.

\begin{figure}
  \resizebox{6cm}{!}{\includegraphics*{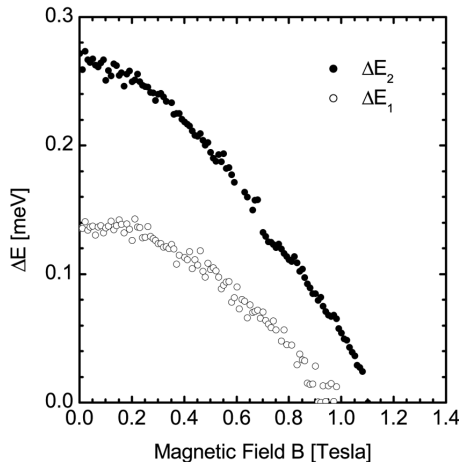}}
  \caption{Singlet-triplet gap $J=\Delta E$ as a function of magnetic
    field for two different gate voltages, as described in the text.
    $\Delta E_1$ and $\Delta E_2$ are also denoted in
    Fig.~\protect\ref{fig:exp.sing-trip}.}
  \label{fig:exp.st-gap}
\end{figure}

For the lowest set of curves in Fig.~\ref{fig:exp.sing-trip},
transport proceeds through the addition and subtraction of a second
electron from a one-electron droplet.  At the lowest curve, transport
is predominately through the ground state of the two-electron droplet
(a spin singlet at low fields); beginning at the curve immediately
above this one, transport through the first excited state (a spin
triplet at low fields) is also allowed.  Hence, the exchange constant
can be directly obtained experimentally from these curves by suitably
calibrating the parameters relating gate voltage to energy.  The
singlet-triplet transition is clearly seen (\textit{cf.}\ etched
vertical
dots\cite{kouwen97:excit.spect.circul,wiel98:singl.tripl.trans}) at a
field of $B_{c1}\approx 0.92$~T.  The upper set of curves corresponds
to adding and removing a third electron from the two-electron system.
After the third electron has left the dot, the resulting two-electron
droplet can either be in the ground state or an excited state.
(Transport through the ground state is the topmost curve.)  Therefore,
it should be possible to extract the same exchange constant from these
curves as described above for the lower set of curves.  Indeed, the
singlet-triplet transition is again clearly seen, but now occurs at a
field of $B_{c2}\approx 1.1$~T.  The singlet-triplet gaps for the two
different cases are shown in Fig.~\ref{fig:exp.st-gap}---the central
experimental result of this work.

Two important conclusions can be drawn from Fig.~\ref{fig:exp.st-gap},
each of which is verified in the subsequent sections.  First, because
the gaps do not close linearly, the confinement potential cannot be
parabolic.  Second, because the two curves do not fall on each other,
the actual shape of the dot must be different for the two curves.
This change can only be due to the gate voltage, and it therefore
follows that \emph{the gates themselves can be used to tune through
  the singlet-triplet transition} and hence tune the ground-state
entanglement of the system.  Figure~\ref{fig:exp.sing-trip} shows how
this may be accomplished.  The solid line demarcates the boundary
between the singlet and triplet ground-state phases. At a fixed field
of 1~T (marked on the figure as a dashed line) the ground-state
entanglement can be tuned to be either a singlet or a triplet solely
by adjusting the gate voltage appropriately.

In the following, we present the theoretical justification of the
above statements.

\section{Theoretical results}
\label{sec:theoretical-results}

We begin this section with a description of the model we shall use
throughout the paper.  We shall work primarily in the 2D harmonic
oscillator (Fock-Darwin) basis, characterized by the two oscillator
quantum numbers $m, n = 0, 1, 2, \dots$ and the spin quantum number
$\sigma=\pm1/2$.  This is the diagonal basis of 2D electrons (taken to
lie in the $x$-$y$ plane) with charge $-e$ and effective mass $m^*$,
moving in a uniform magnetic field $\vect{B}=(0,0,B)$ oriented
perpendicular to the 2DEG plane, and with a parabolic confinement
potential $V_{\text{par}} = m^*\omega_0^2(x^2+y^2)/2$.  The
single-particle energy levels are given by the familiar Zeeman-split
Fock-Darwin spectrum:\cite{fock28,darwin30,jacak97:quant.dots}
\begin{equation}
  \label{eq:fd-spect}
  \varepsilon_{mn\sigma} = \Omega_+ \left(n+\frac{1}{2}\right) 
  + \Omega_- \left(m+\frac{1}{2}\right) + g\mu_BB\sigma.
\end{equation}
The first two terms in this equation are the oscillator energies, with
$\Omega_\pm = (\sqrt{\omega_c^2 + (2\omega_0)^2} \pm \omega_c)/2$,
$\omega_c = eB/(m^*c)$ denoting the cyclotron frequency, and
$\omega_0$ the parabolic confinement frequency.  The final term is the
Zeeman energy, where $g\mu_BB \approx 0.012 \hbar \omega_c$ in GaAs.

Neglecting environmental influences, the Hamiltonian of an isolated
quantum dot can be written in the Fock-Darwin
basis\cite{jacak97:quant.dots} as
\begin{multline}
  \label{eq:hamil.full}
  H = \sum_{i,\sigma} \varepsilon_{i\sigma} 
  c^\dag_{i\sigma} c_{i\sigma} + 
  \gamma \sum_{i,j,\sigma} h_{ij} c^\dag_{i\sigma} c_{j\sigma}
  \\ \mbox{} +
  \alpha \!\sum_{\substack{i,j,k,l,\\ \sigma,\sigma'}} V_{ij}^{k\ell}
  c^\dag_{i\sigma}c^\dag_{j\sigma'}c_{k\sigma'}c_{\ell\sigma},
\end{multline}
where the Latin indices $i,j,k,l$ are a composite denoting the two
oscillator quantum numbers $m$ and $n$.  The operator
$c^\dag_{i\sigma} \rightarrow c^\dag_{mn\sigma}$ creates a particle in
the state $|mn\sigma\rangle$ with the $z$-component of angular
momentum $(m-n)$, and $c_{i\sigma}$ is the conjugate annihilation
operator.  Unless otherwise noted, all energies are expressed in units
of the effective Rydberg $\text{Ry} = m^*e^4/(2\hbar^2)$ ($\approx
5.9$~meV in GaAs), and all lengths in units of the effective Bohr
radius $a_0=\hbar^2/(m^*e^2)$ ($\approx 9.8$~nm in GaAs).

The diagonal one-body term---the first term in
Eq.~(\ref{eq:hamil.full})---is just the Fock-Darwin energy coming from
parabolic confinement; $\varepsilon_{i\sigma} \rightarrow
\varepsilon_{mn\sigma}$ is given in Eq.~(\ref{eq:fd-spect}).  We
consider the \emph{total} confinement to be composed of a parabolic
piece plus a non-parabolic piece; the parabolic piece, along with the
kinetic energy, is incorporated into the diagonal term; the
non-parabolic piece is represented by the off-diagonal one-body
term---the second term in Eq.~(\ref{eq:hamil.full})---whose overall
strength is governed by the dimensionless parameter $\gamma$.  This
term directly affects the single-particle spectrum, and also
significantly alters the singlet-triplet transition in the
two-electron droplet.  We discuss this term in detail in
Sec.~\ref{sec:non-parab-conf}.

Finally, the two-body term in Eq.~(\ref{eq:hamil.full}) represents
interactions, where the matrix element $V_{ij}^{kl} \rightarrow
\langle m_1,n_1;m_2,n_2 | e^2 /(\epsilon |\vec{r}_1 - \vec{r}_2|) |
m_3, n_3; m_4, n_4 \rangle$ is the full Coulomb interaction in the 2D
harmonic oscillator basis ($\epsilon$ is the dielectric constant); an
exact expression is given by~\cite{hawry93:far.infrar.absor}
\begin{widetext}
  \begin{multline}
    \label{eq:v-coul}
    V_{ij}^{kl} \rightarrow \frac{E_0}{\sqrt{2 \pi}} \,
    \delta_{R_L,R_R} \frac{(-1)^{n_2+m_2+n_3+m_3}}
    {\sqrt{n_1!m_1!n_2!m_2!n_3!m_3!n_4!m_4!}}
    \sum_{k_1=0}^{\text{min}(m_1,m_4)} k_1!
    \begin{pmatrix} m_1\\k_1 \end{pmatrix}
    \begin{pmatrix} m_4\\k_1 \end{pmatrix}
    \sum_{k_2=0}^{\text{min}(n_1,n_4)} k_2! 
    \begin{pmatrix} n_1\\k_2 \end{pmatrix}
    \begin{pmatrix} n_4\\k_2 \end{pmatrix} \\ \mbox{} \times
    \sum_{k_3=0}^{\text{min}(m_2,m_3)} k_1! 
    \begin{pmatrix} m_2\\k_3 \end{pmatrix}
    \begin{pmatrix} m_3\\k_3 \end{pmatrix}
    \sum_{k_4=0}^{\text{min}(n_2,n_3)} k_2! 
    \begin{pmatrix} n_2\\k_4 \end{pmatrix}
    \begin{pmatrix} n_3\\k_4 \end{pmatrix}
    \left(\frac{-1}{2}\right)^k \Gamma\left(k+\frac{1}{2}\right).
  \end{multline}
\end{widetext}
The energy scale of the Coulomb interaction is set by $E_0 \equiv
\sqrt{\pi/2} \, e^2/(\epsilon\ell_0)$ where the hybrid length
$\ell_0^2 = \hbar c / (eB \sqrt{1+4\omega_0^2/\omega_c^2})$.  $E_0$ is
the sum of all exchange energies in the lowest Landau level,
\textit{i.e.}, $\sum_{m=0}^\infty \langle m,0;0,0 | e^2 /(\epsilon
|\vec{r}_1 - \vec{r}_2|) | m,0;0,0 \rangle = E_0$, and, additionally,
$\langle 0,0;0,0 | e^2 /(\epsilon |\vec{r}_1 - \vec{r}_2|) | 0,0;0,0
\rangle = E_0/\sqrt{2}$.  The Coulomb interaction conserves angular
momentum $R \equiv \sum_i (m_i-n_i)$.  This is enforced in
Eq.~(\ref{eq:v-coul}) by the Kronecker delta function: $R_L =
(m_1-n_1) + (m_2-n_2)$, $R_R = (m_3-n_3) + (m_4-n_4)$.  Also in
Eq.~(\ref{eq:v-coul}), $k=(m_1+m_2+n_3+n_4)-(k_1+k_2+k_3+k_4),\ 
\text{and}\ \Gamma(k+1/2)$ is the Gamma function.

The dimensionless parameter $\alpha$ in Eq.~(\ref{eq:hamil.full})
controls the strength of the Coulomb interaction, with $\alpha=1$
representing ``bare'' Coulomb interactions.  At long length scales,
screening effects from the nearby metallic gates and leads decrease
the strength of the Coulomb interaction.  At short length scales, the
finite width of the 2DEG layer also decreases the strength of
interactions.  Since Coulomb interactions are not the primary focus at
present, we shall use the parameter $0 \le \alpha \le 1$ to describe
the strength of Coulomb interactions, rather than considering a more
sophisticated functional form.

In order to determine which of the main results are specific to the
details of the confinement and which are more general, we shall first
consider the usual parabolic confinement, and subsequently investigate
the particular deviations from parabolicity present in our device.

\subsection{Parabolic confinement}
\label{sec:inter-parab-case}

In this section, we consider the case of pure parabolic confinement
($\gamma=0$).  We shall see that our central result of voltage-tuned
entanglement is already present in this simple case.

\subsubsection{Non-interacting electrons}

\begin{figure}
  \resizebox{8cm}{!}{\includegraphics*{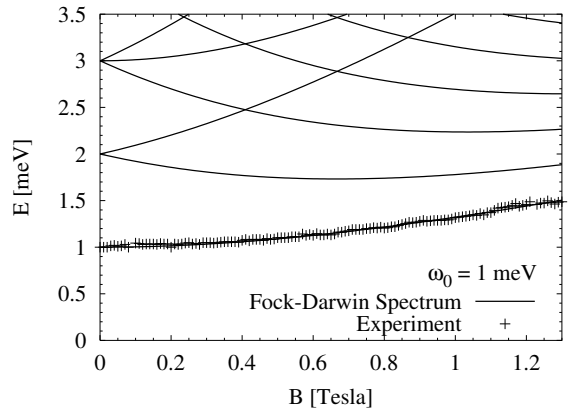}}
  \caption{Comparison of experiment and theory for the one-electron
    droplet.  The data was fit to Eq.~(\ref{eq:fd-spect}), with
    $\omega_0$ as a fitting parameter.}
  \label{fig:exp-theory-1e}
\end{figure}

The single-particle problem with parabolic confinement
($\gamma=\alpha=0$) yields the Fock-Darwin spectrum in
Eq.~(\ref{eq:fd-spect}).  This approximation should be most valid for
the lowest-energy level of the one-electron droplet; in addition to
having no intra-dot Coulomb interactions, the zero-point energy should
be smallest for the one-electron droplet.  A comparison of the
Fock-Darwin spectrum and experiment is shown in
Fig.~\ref{fig:exp-theory-1e}.
The experimental points are the position of current peak as a function
of magnetic field for the one-electron droplet.  The parabolic
confinement frequency $\omega_0$ was used as a fitting parameter, with
$\omega_0=1$~meV being displayed in the figure.  Although the data is
well-fit for this value, we shall find a rather different situation
for the two-electron droplet.

\subsubsection{Interacting electrons}

When Coulomb interactions are switched on (but with $\gamma=0$ for the
moment), $m$ and $n$ are no longer good quantum numbers, but, since
circular symmetry is still manifest, the total angular momentum $R$ is
indeed conserved, as are total spin $S$ and total $S_z$.  The
Hamiltonian can therefore be diagonalized in each $(R,S,S_z)$ subspace
separately.

We have numerically diagonalized Eq.~(\ref{eq:hamil.full}) with
$\gamma=0$ according to the following procedure: We work in a fixed
$(R,S_z)$ subspace---alternatively,\cite{amjp.unpub} one may work in a
fixed $(R,S,S_z)$ subspace, a particularly useful approach for larger
particle numbers---and we use the 2D harmonic oscillator basis, with
the Coulomb matrix elements given by Eq.~(\ref{eq:v-coul}).  We
truncate the infinite-dimensional Hilbert space by introducing a
high-energy cutoff $E_{\text{cutoff}}$; For each $N$-particle basis
vector $|m_1n_1\sigma_1,\ldots,m_Nn_N\sigma_N\rangle$, we calculate
its ($\alpha=0$) eigenenergy and discard it if this energy is greater
than $E_{\text{cutoff}}$.  We then numerically diagonalize the
resulting finite-dimensional Hamiltonian $\langle
m_Nn_N\sigma_N,\ldots,m_1n_1\sigma_1 | H |
m_1'n_1'\sigma_1',\ldots,m_N'n_N'\sigma_N' \rangle$ (with finite
$\alpha$) to obtain both the eigenstates and the spectrum.  We then
keep repeating this process with a progressively larger
$E_{\text{cutoff}}$ until the eigenvalues converge to a constant
value.

\begin{figure}
  \resizebox{8cm}{!}{\includegraphics*{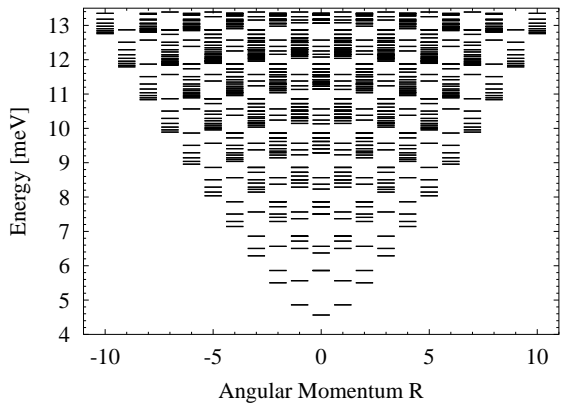}}
  \caption{Eigenvalues of the interacting two-electron droplet with
    parabolic confinement.  The plot is for $B=0$~T, $\omega_0=1$~meV,
    and for singlets and triplets with $S_z = 0$.}
  \label{fig:par-eigs}
\end{figure}

For small magnetic fields ($B < 5$~T, for $\omega_0=1$~meV) and two
electrons, convergence is reached rather quickly.  For example, for a
112-dimensional Hilbert space, convergence to within 4\% has been
achieved for the lowest 65 eigenstates for $(N,B,R,S_z) = (2,0,1,0)$
(this spin value includes both singlet and triplet states), and to
within 0.5\% for the lowest 49 eigenstates.  The eigenvalues for
$R=-10$ to 10, $B=0$, and $S_z=0$ are shown in
Fig.~\ref{fig:par-eigs}.

\begin{figure}
  \resizebox{8cm}{!}{\includegraphics*{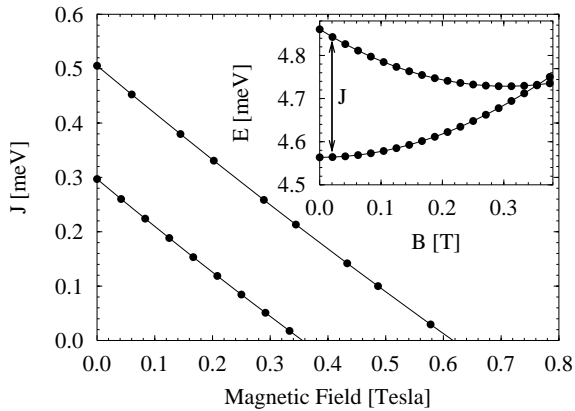}}
  \caption{Inset:  Lowest-energy singlet $|S,S_z\rangle = |0,0\rangle$
    (lower curve) and triplet $|1,-1\rangle$ (upper curve) for the
    two-electron droplet with $\alpha=1$ and $\omega_0=1$~meV.  The
    singlet-triplet gap $J$ is indicated.  Main plot: Singlet-triplet
    gap as a function of magnetic field for the two-electron droplet
    with $\alpha=1$.  The upper curve has $\omega_0=1.5$~meV and the
    lower has $\omega_0=1$~meV.}
  \label{fig:parab-sing-trip}
\end{figure}

Experimentally, we have seen that the magneto-transport data of the
one-electron droplet is very well described by parabolic confinement
with a confinement frequency of $\omega_0=1$~meV.  In the two-electron
droplet, where Coulomb interactions are now relevant, the ground-state
singlet-triplet transition is experimentally seen to occur at
approximately $B_c = 1$~T.  If we assume the confinement frequency
remains constant at 1 meV, then $B_c = 1$~T occurs for $\alpha\approx
0.2$, and thus Coulomb interactions are significantly reduced from
their bare value.  Alternatively, since the critical field scales with
the ratio $\omega_0/(\alpha E_0)$, $\omega_0$ may increase, rather
that $\alpha$ decrease, to give the same effect.  This is shown in
Fig.~\ref{fig:parab-sing-trip}, where all curves are for the case
$\alpha=1$.
The main plot shows the evolution of the singlet-triplet gap $J$ with
magnetic field.  The lower curve has $\omega_0=1$~meV while the upper
has $\omega_0=1.5$~meV.  The inset shows the actual singlet-triplet
crossing for $\omega_0=1$~meV.  This simple model of parabolic
confinement with full Coulomb interactions is clearly insufficient to
quantitatively reproduce the experimental findings of
Fig.~\ref{fig:exp.st-gap}.  The linear closing of the gap in the
theory---there are actually very slight deviations from linearity not
discernible in Fig.~\ref{fig:parab-sing-trip}---appears to be a
feature of Coulomb interactions combined with parabolic confinement.
The experimental curves of Fig.~\ref{fig:exp.st-gap} are thus an
indication that the confinement is non-parabolic.

What this simple theory \emph{does} capture, is that the
critical-field $B_c(\omega_0)$ is controlled by the gate voltage via
its influence on the confining frequency $\omega_0(V_g)$ for a
\emph{fixed} particle number $N$.  Thus, already at this level, we see
how the gate voltage, at fixed uniform magnetic field, can be used to
tune through the singlet-triplet transition.

The parabolic-confinement model is insufficient to reproduce both the
critical field $B_c$ and the zero-field gap $J_0=J(B=0)$.
Finite-width effects ($\alpha<1$) or an increasing $\omega_0$ serve to
increase both $B_c$ and $J_0$.  In the following section, we
investigate the influence of non-parabolic confinement on the
singlet-triplet gap.

\subsection{Non-parabolic confinement}
\label{sec:non-parab-conf}

The confinement potential produced by the gate geometry shown in
Fig.~\ref{fig:gates} exhibits only approximate circular symmetry and
this explicit symmetry breaking can be clearly seen in experiment
already at the two-electron level, as shown in
Fig.~\ref{fig:exp.st-gap}.  This deviation from parabolicity is
included as the second term of our model in Eq.~(\ref{eq:hamil.full}),
and it is the influence of this term we investigate in this section.

\begin{figure}
  \resizebox{8cm}{!}{\includegraphics*{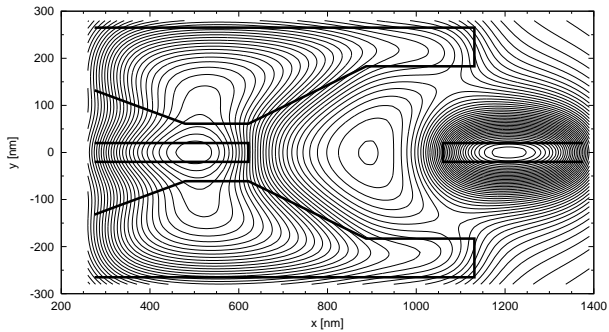}}
  \caption{Contour plot (thin lines) of the
    two-dimensional confinement potential produced by the gates (thick
    lines) located 90~nm above the 2DEG.}
  \label{fig:poten.prof} 
\end{figure}

The first problem is to obtain a functional form for the confinement
potential.  Since the gate voltages are the primary contribution to
the confinement potential, the simplest approach is to consider the
electrostatic potential in the 2DEG induced by the gate voltages.
Defining the $x$-$y$ plane to be the plane of the gates so that the
potential $V(\vect{r},z=0)$ is experimentally given ($\vect{r} \equiv
(x,y)$), an analytic expression can be
derived\cite{davies95:model.patter.two} for the potential $V(\vect{r},
z)$ at an arbitrary point:
\begin{equation}
  \label{eq:poten.anal}
  V \left( \vect{r}, z \right) = \int \frac{d\vect{r}'}{2\pi} \,
  |z| \, \frac{V(\vect{r}', 0)}{
    \left( z^2 + \left| \vect{r}-\vect{r}' \right|^2 \right)^{3/2} }.
\end{equation}
This equation yields the correct $z \rightarrow 0$ limit, as well as
$\partial V / \partial z \rightarrow 0$ for $|z| \rightarrow \infty$.
The integration, performed at each point $\vect{r}$ in the 2DEG plane
($z=90\ \text{nm}$), yields the potential which laterally confines
electrons.  A contour plot of this confinement is shown in
Fig.~\ref{fig:poten.prof}.
We have neglected the contribution of any $B$-dependant effects of the
edge-states (i.e., the leads).  In the Coulomb-blockade regime we are
interested in here, the $B$-dependence of the lead states will
primarily influence the tunneling rates into and out of the dot.  That
is to say, the amplitude of the current will be affected, rather than
the spectrum of the dot.

The confinement potential in Fig.~\ref{fig:poten.prof} can be viewed
as a sum of a parabolic dot and a parabolically-confined semicircular
wire of diameter $D$ which intersects the quantum point contacts (seen
as saddle-points in Fig.~\ref{fig:poten.prof}) and the center of the
dot.  These considerations lead to an analytic expression which very
closely approximates the (numerically-derived) potential in
Fig.~\ref{fig:poten.prof}.  This potential is given by $H_{\text{par}}
+ \gamma H_{\text{nopar}}$, where $H_{\text{par}} = (1/2) m^\ast
\omega_0^2 (x^2+y^2)$ is the usual parabolic confinement, and
\begin{equation}
  \label{eq:Hpot}
  H_{\text{nopar}} = \frac{1}{2} m^\ast \omega_0^2
  \left( x - \frac{y^2}{D} \right)^2
\end{equation}
is the non-parabolic piece of the confinement potential.  In
Eq.~(\ref{eq:hamil.full}), the parabolic piece is incorporated into
the diagonal one-body term, while the off-diagonal one-body term is
the second-quantized version of Eq.~(\ref{eq:Hpot}) with
$h_{ij}\rightarrow \left\langle m n | H_{\text{nopar}} | m' n' \right
\rangle$.

The computational consequences of the explicit symmetry-breaking terms
($\gamma \neq 0$) are that the Hilbert-space truncation scheme must be
altered to incorporate the mixing of different angular-momentum
subspaces.  One possibility for the interacting problem is to begin
with the single-particle states (no longer simple Fock-Darwin states)
and then solve the interacting problem in the exact single-particle
basis.  Another approach, indeed the one we present below, is to treat
the parabolically-confined interacting problem ($\gamma=0,\ \alpha\neq
0$) exactly, as was done in Sec.~\ref{sec:inter-parab-case}, and, in
this basis, treat the one-body symmetry-breaking terms.

That Eq.~(\ref{eq:Hpot}) mixes different $R$ subspaces can be
explicitly seen by re-writing the position operators in terms of the
usual oscillator ladder operators:\cite{jacak97:quant.dots} $x =
\ell_0 (a^\dag + a + b^\dag + b)/\sqrt{2}$ and $y = -i\ell_0 (a^\dag -
a - b^\dag + b)/\sqrt{2}$, where $a^\dag = \sum (n+1)
c^\dag_{m,n+1,\sigma} c_{m,n,\sigma}$ and $b^\dag = \sum (m+1)
c^\dag_{m+1,n,\sigma} c_{m,n,\sigma}$.  Using these relations, we
rewrite the second term of Eq.~(\ref{eq:hamil.full}) as
\begin{subequations}
  \label{eq:hpot-bose}
  \begin{equation}
    \gamma \sum_{i,j,\sigma} h_{ij} c^\dag_{i\sigma} c_{j\sigma} = 
    \frac{\gamma \omega_0^2/4}{\sqrt{\omega_c^2+4\omega_0^2\,}} 
    \sum_{\delta R=0}^4
    \left(\delta V_{\delta R} + \delta V_{\delta R}^\dag \right),
  \end{equation}
  where $\delta V_{\delta R}$ changes the angular momentum ($m-n$) of
  the single-particle state $|mn\rangle$ by an amount $\delta R$,
  $\delta V_{\delta R}^\dag = \delta V_{-\delta R}$ ($\delta V_0$ is
  Hermitian), and where
  \begin{align}
    \delta V_0& = 3 \beta^2 ( a^{\dag 2} d^2 + 2 a^\dag d^2 b
                  + d^2 b^2 ) \\
              & \quad + (1+12\beta^2) ( a^\dag d + d b )
                  + (1+6\beta^2), \notag \\
    \delta V_1& = -2\beta ( a^\dag d^2 + 2d + d^2 b ), \\
    \delta V_2& = -4\beta^2 ( a^\dag d^3 + d^3 b )
                  + (1-12\beta^2) d^2, \\
    \delta V_3& = 2\beta d^3, \\ \intertext{and}
    \delta V_4& = \beta^2 d^4,
  \end{align}
\end{subequations}
with $d=(a+b^\dag)$, and $\beta = D^{-1} (\omega_c^2 +
4\omega_0^2)^{-1/4}$.

Before going on to the main task of investigating the singlet-triplet
transition in this non-parabolic potential, we first investigate how
the Fock-Darwin spectrum is affected by these symmetry-breaking terms.

\subsubsection{Non-interacting electrons}
\label{sec:single-part-levels}

We have been unable to find an exact analytic solution to the
non-parabolically confined ($\gamma\neq0$) single-particle
($\alpha=0$) problem, and we therefore employ a numerical treatment.
It is simplest to again work in the 2D harmonic oscillator
(Fock-Darwin) basis.  In the single-particle problem, Zeeman effects
are rather trivial and shall therefore be neglected in the present
discussion.  We include a fixed number of Fock-Darwin states in our
Hilbert space, diagonalize the Hamiltonian, and repeat with a larger
number of Fock-Darwin states, progressively increasing the
Hilbert-space dimension until convergence of the spectrum is attained
for the lowest few levels.

\begin{figure}
  \resizebox{8cm}{!}{\includegraphics*{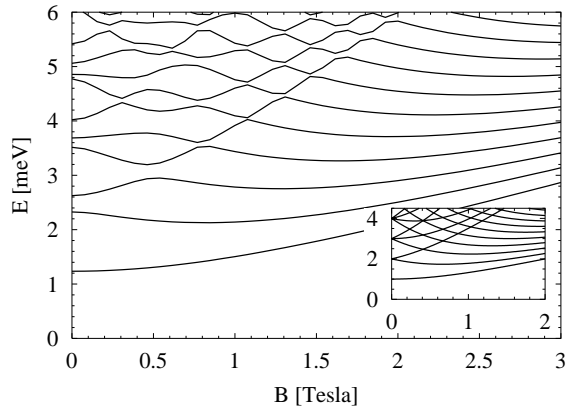}}
  \caption{Main plot: Single-particle spectrum as a function of
    magnetic field for a non-parabolically confined dot with
    $\omega_0=1$~meV, $\gamma=1$, and $D=10a_0$.  Remnants of
    shell-structure at zero field can be discerned.  Inset: The same
    spectrum, but with $\gamma=0$, i.e., the Fock-Darwin spectrum.}
  \label{fig:one-body-deformed}
\end{figure}

An example of the resulting spectrum is shown in
Fig.~\ref{fig:one-body-deformed}, with the equivalent Fock-Darwin
spectrum as an inset.
The most dramatic effect of non-parabolicity is at low magnetic
fields, where the shell structure is heavily renormalized, although
its remnants are still observable.  The plot was computed using the
1891 lowest-energy Fock-Darwin levels---corresponding, at zero field,
to the first 61 shells in the parabolic case---and with
$\omega_0=1$~meV, $\gamma=1$, and $D=10a_0$.  At much larger $\gamma$,
the shell-splitting becomes so large that different shells overlap,
and a Fock-Darwin description at low fields becomes dubious.  Apparent
anti-crossings are also seen in Fig.~\ref{fig:one-body-deformed},
whereas the Fock-Darwin spectrum contains only crossings.  Another
important feature is that the $\nu=2$ line at moderate fields is still
clearly visible, even for larger $\gamma$; beyond this point, field
effects begin to play a more prominent role than non-parabolicity
effects, whose presence in the spectra becomes concealed.  Thus, we do
not expect non-parabolicity effects to play an appreciable role beyond
$\nu=2$.\cite{ciorg02:collap.spin.singl}

\subsubsection{Interacting electrons}
\label{sec:interact-prob}

In section~\ref{sec:inter-parab-case}, we treated the
parabolically-confined interacting case by first solving for the
non-interacting (Fock-Darwin) case; these states were then used as the
basis states in which the interacting problem was solved.  Continuing
the progression, we now use the exact \emph{interacting} many-body
states computed in Sec.~\ref{sec:inter-parab-case} as the basis states
in which we treat the non-parabolic piece of the confinement
potential.

The method of truncating the Hilbert-space must again be chosen.  As
an example, Fig.~\ref{fig:par-eigs}---for all angular momenta since
they are no longer conserved---may be considered the Hilbert-space for
the particular case of $\omega_0=1$~meV, $B=0$~T, and $S_z=0$.  Two
methods of truncating this Hilbert-space can be considered.  First,
the $k$ lowest-energy states within each angular momentum channel may
be chosen as the reduced Hilbert-space, with all higher-energy states
discarded.  For example, if $k=10$ and there are 21 angular momentum
channels (as in Fig.~\ref{fig:par-eigs}), then the Hilbert space has
$10 \times 21 = 210$ dimensions.  In this scheme, the cutoff energy is
variable, but the number of states within each angular momentum
channel is fixed.  The second method is to employ a fixed energy
cutoff $E_{\text{cutoff}}$ and to allow a variable number of states
within each angular momentum channel; all states above
$E_{\text{cutoff}}$, regardless of angular momentum, are discarded and
all states below $E_{\text{cutoff}}$, regardless of angular momentum,
are retained.  In principle, it matters little which truncation scheme
is used so long as each method, of course, converges to the same
values.  In practice, the second method, with a fixed cutoff energy,
achieves convergence faster.

\begin{figure}
  \resizebox{8cm}{!}{\includegraphics*{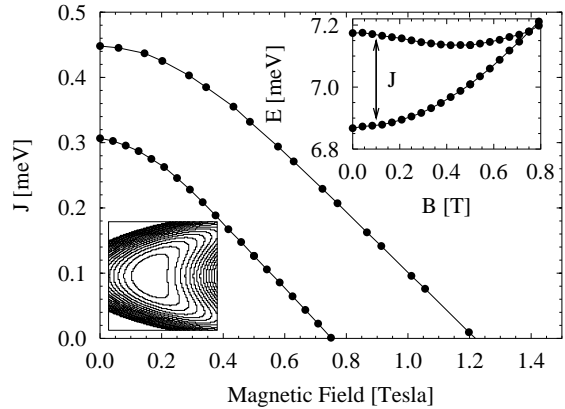}}
  \caption{Main Plot: Singlet-triplet gap $J$ as a function of
    magnetic field for $\omega_0=1$~meV (lower curve) and
    $\omega_0=1.5$~meV (upper curve) for non-parabolically confined
    interacting electrons. Upper Inset: Lowest-energy singlet
    $|S,S_z\rangle = |0,0\rangle$ (lower curve) and triplet
    $|1,-1\rangle$ (upper curve) as a function of magnetic field for
    $\omega_0=1$~meV.  Lower Inset: Shape of total confinement
    potential in real space.  All curves in the main plot and in the
    insets are for $\alpha=1$, $\gamma=3$, and $D=5 a_0$.  This figure
    is the analog to Fig.~\ref{fig:parab-sing-trip} for non-parabolic
    confinement.}
  \label{fig:sing-trip}
\end{figure}

Figure~\ref{fig:sing-trip} shows results analogous to
Fig.~\ref{fig:parab-sing-trip}, but for the non-parabolic confinement
discussed above.
All curves shown in Fig.~\ref{fig:sing-trip} have $(\alpha, \gamma, D)
= (1, 3, 5a_0)$.  In the main plot, the gap $J$ is plotted for
$\omega_0 =1$~meV (lower curve) and $\omega_0=1.5$~meV (upper curve).
The upper inset shows the singlet-triplet transition for
$\omega_0=1$~meV, and the lower inset shows a sketch of the shape of
the confinement potential for these parameter values.  All these
parameters, except for $\gamma$, are as in
Fig.~\ref{fig:parab-sing-trip}.  The non-parabolic spectrum differs
significantly from the parabolic spectrum, particularly at small
field, where the triplet has a much weaker field dependence in the
present case relative to the parabolic-confinement case.  This is the
behavior also seen in experiment.  (See Fig.~\ref{fig:exp.sing-trip}.)

In general, the non-parabolic model yields results much closer to
experiment than parabolic confinement, and it is clear that the
confinement potential in experiment is not parabolic.  The available
phase space---with variation in $\omega_0$, $\alpha$, $\gamma$, and
$D$---is rather large and so an optimal fit has not been performed.
Nevertheless, the theoretical curves in Fig.~\ref{fig:sing-trip} are
in qualitative agreement with the experimental curves in
Fig.~\ref{fig:exp.st-gap}.  Future analyses of high source-drain
spectroscopy with higher electron numbers will produce additional
constraints which will more meaningfully reduce the range of these
parameters.

\begin{figure}
  \resizebox{8cm}{!}{\includegraphics*{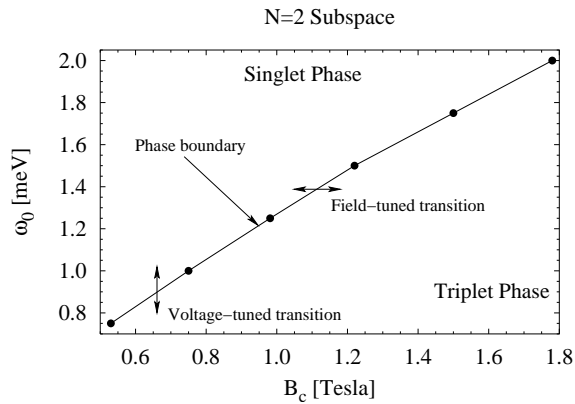}}
  \caption{Spin phase diagram for the two-electron system in the
    $\omega_0$-$B_c$ plane.  This plot has $\alpha=1$, $\gamma=3$, and
    $D=5a_0$, and $\omega_0$ is controlled by the gate voltages.  The
    usual field-induced singlet-triplet transition and the
    voltage-induced transitions are indicated.}
  \label{fig:phase-diag}
\end{figure}

The central point is that the particular critical field $B_c$ obtained
is \emph{itself} a function of the confining-potential parameters,
$\omega_0$, $\gamma$, and $D$.  Since each of these parameters is
controlled by the voltage on the gates shown in Fig.~\ref{fig:gates},
it follows that \emph{the gates themselves can be used to tune through
  the singlet-triplet transition} and hence tune the ground-state
entanglement of the system.  This is shown explicitly in
Fig.~\ref{fig:phase-diag} where we plot the spin phase diagram in the
$\omega_0$-$B_c$ plane.
In this figure, the singlet-triplet gap $J$ is computed for various
values of $\omega_0$ (tunable by the gate voltage); the critical field
$B_c$ is then extracted from the solution to $J (\omega_0, B_c) \equiv
0$.  In practice, the range of $\omega_0$ which is experimentally
accessible is delimited by the $N=3$ and $N=1$ subspaces as the gate
voltage is swept.  However, if the accessible range is sufficient to
be seen by experiment, a voltage-tuned singlet-triplet transition at
fixed $B$ is achievable.  As shown in
Sec.~\ref{sec:experimental-results}, this is indeed the case for the
present experiment.

\section{Discussion}
\label{sec:conclusion}

The fact that the gate voltage (or set of gate voltages in the present
case) controls not only the chemical potential of the dot but also the
\emph{shape} of the dot has been exploited in the present work to tune
the ground-state entanglement of an electron pair in a lateral quantum
dot.  The experimental evidence clearly shows that the two-electron
singlet-triplet transition occurs at a critical field which depends on
the gate voltage.  The confinement potential is not parabolic (nor
elliptical) and this allows great flexibility in changing the shape of
the dot while simultaneously keeping the number of confined electrons
fixed.  However, the experimental demonstration of voltage-tuned
entanglement is not dependent upon the precise shape of the potential,
and should be achievable in a wide range of potential shapes.  The
main requirement is that the gate voltages appreciably change the
shape of the potential while the particle number remains constant.
The present is a modest step towards the construction of a quantum
gate in a solid-state environment; it does not, for example,
demonstrate the \emph{unitary} evolution of the system between singlet
and triplet states.

Because the tunneling barriers into and out of the dot are large, only
the leading order contribution to the tunneling current can be seen in
our experiment.  In principle, however, the tunnel barriers can be
reduced while still remaining in the Coulomb blockade regime in order
to measure an appreciable cotunneling
current.\cite{defran01:elect.cotun} In this way, the singlet-triplet
transition may be experimentally probed throughout the $N=2$ subspace,
and thus give a more stringent test of theory.

In a double-dot system with one trapped electron apiece, essentially
the same spin physics occurs, and thus a voltage-tuned ground-state
transition should also occur along with the consequent implications
for quantum computing.  Alternatively, the current work may be
speculatively viewed as a possible gate-controlled single-qubit
operation, where the single coded qubit exists in a single quantum dot
containing two (or more) electrons.  This conjecture will be more
fully developed in a future publication.

\begin{acknowledgments}
  A. S. S., P. H., and J. K. acknowledge the support of the
  Nanoelectronics program of the Canadian Institute for Advanced
  Research.
\end{acknowledgments}

\bibliography{bibabbrevs,0D,misc,./sing-trip}

\end{document}